\documentclass[twocolumn,twocolappendix]{aastex631} 

\graphicspath{{./}{fig/}}

\shorttitle{CHIMERA KBO Occultation Constraints}
\shortauthors{Zhang et al.}

\journalinfo{The Astronomical Journal (preprint)} 

\received{2023 September 16}
\revised{2023 October 10}
\accepted{2023 October 11}

\renewcommand{\edit}[2]{#2} 

\begin{document}

\title{CHIMERA Occultation Constraints on the Abundance of Kilometer-scale Kuiper Belt Objects}

\author[0000-0002-6702-191X]{Qicheng Zhang}
\affiliation{Division of Geological and Planetary Sciences, California Institute of Technology, Pasadena, CA 91125, USA}

\author[0000-0002-7083-4049]{Gregg W. Hallinan}
\affiliation{Department of Astronomy, California Institute of Technology, Pasadena, CA 91125, USA}

\author{Navtej S. Saini}
\affiliation{Jet Propulsion Laboratory, California Institute of Technology, Pasadena CA 91109, USA}
\affiliation{Department of Astronomy, California Institute of Technology, Pasadena, CA 91125, USA}

\author[0000-0002-0298-8089]{Hilke E. Schlichting}
\affiliation{Department of Earth, Planetary, and Space Sciences, University of California, Los Angeles, CA 90095, USA}

\author{Leon K. Harding}
\affiliation{Northrop Grumman Space Systems, Redondo Beach, CA 90278, USA}
\affiliation{Department of Astronomy, California Institute of Technology, Pasadena, CA 91125, USA}

\author{Jennifer W. Milburn}
\affiliation{Department of Astronomy, California Institute of Technology, Pasadena, CA 91125, USA}

\correspondingauthor{Qicheng Zhang}
\email{qicheng@cometary.org}

\begin{abstract}
Occultations provide indirect sensitivity to the number density of small Kuiper Belt objects (KBOs) too faint to directly detect telescopically. We present results from the Caltech HI-speed Multicolor camERA (CHIMERA) survey with the Palomar Hale Telescope, which monitored stars over the central $5'\times5'$ of the M22 globular cluster along the ecliptic plane for serendipitous occultations by kilometer-scale KBOs over 63~hr across 24 nights at a 33~Hz frame rate simultaneously in $i'$ and $g'$. We adapted dense-field photometry and occultation template fitting techniques to this dataset, finding a 95\% confidence upper limit on the occultation rate corresponding to an ecliptic sky density of ${\lesssim}10^7$~deg$^{-2}$ of $>$1~km diameter classical KBOs. We discuss a few of the occultation-like light curve signatures at the edge of the sensitivity limit responsible for setting the upper bounds, and their likely nonviability as true occultations.
\end{abstract}


\section{Introduction}

Kuiper Belt objects (KBOs) populate the outer solar system beyond the orbit of Neptune \citep{jewitt1993,jewitt1999}. These objects provide a record of the evolution of the early solar system \citep[e.g.,][]{morbidelli2003,kenyon2004,schlichting2013} and supply the transient population of Jupiter-family comets (JFCs) in the inner solar system \citep[e.g.,][]{duncan1997}. However, while objects larger than several kilometers in diameter can be directly found and observed by their reflected sunlight \citep[\edit{2}{e.g.},][]{millis2002,fraser2008,fuentes2008,bernardinelli2020}, the presumably more abundant kilometer-scale and smaller KBOs are too faint to be observed in this way with modern survey instrumentation.

The population of these small KBOs can thus currently only be probed indirectly, such as by monitoring for serendipitous occultations of background stars by such objects \citep{bailey1976}---events generally lasting less than a second thus requiring high-cadence photometry to reliably detect---and several occultation surveys have been conducted to constrain the number density of small KBOs. \citet{roques2006} used simultaneous high-speed imaging in $g'$ and $i'$---thus obtaining two sets of simultaneous and only partially correlated light curves for each star---through the 4.2~m William Herschel Telescope. They reported three candidate events, but with properties inconsistent with those expected for true KBO occultations. \citet{bianco2009} used single-channel, $r'$ imaging from the 6.5~m MMT Observatory's Megacam and likewise reported no likely KBO occultations. The much deeper Taiwanese--American Occultation Survey \citep[TAOS;][]{alcock2003} used a set of four separate 0.5~m telescopes simultaneously observing the same star field, and similarly set only an upper limit on the small KBO abundance \citep{zhang2008,zhang2013}.

Another search by \citet{schlichting2009} and \citet{schlichting2012} of archival data from the 2.4~m Hubble Space Telescope (HST)'s Fine Guidance Sensors (FGS)---observing from above the atmosphere, thus providing photometry free of atmospheric scintillation effects---found two claimed occultations, while a search through Convection, Rotation and planetary Transits (CoRoT) spacecraft imagery by \citet{liu2015} yielded another dozen claimed occultations. Additionally, \citet{arimatsu2019} claimed an occultation detection by their ground-based Organized Autotelescopes for Serendipitous Event Survey (OASES) system of two co-aligned 0.3~m telescopes, although \citet{nir2023} more recently report a nondetection from a survey with their 0.55~m Weizmann Fast Astronomical Survey Telescope \citep[W-FAST;][]{nir2023} in tension with the OASES detection claim.

Curiously, crater counting with New Horizons imagery \citep{singer2019} and extrapolations of direct telescopic surveys \citep{kavelaars2021} have suggested kilometer-scale KBOs may be depleted to orders of magnitude below the abundances implied by the claimed occultations. The difference suggests that either these studies are sampling a different population within the Kuiper Belt with significantly fewer kilometer-sized objects than that sampled by occultations, they made erroneous assumptions in converting their respective data into KBO number densities, or the claimed occultation detections are actually artifacts missed by the respective statistical analyses. This discrepancy motivates further, more sensitive occultation surveys to validate or reject the earlier claims.

In this paper, we present results from a survey carried out with the Caltech HI-speed Multicolor camERA \citep[CHIMERA;][]{harding2016} on the 5.1~m Palomar Hale Telescope over 2015--2017. CHIMERA was specifically designed for this survey, and uses a dichroic and two EMCCD cameras to provide high-speed imaging simultaneously in both a blue channel with a $g'$ filter and a red channel with an $i'$ filter, producing data of a similar nature to, but far more sensitive than, that previously collected by \citet{roques2006}. We discuss the nature of our observations, the photometric reduction process, and the use of occultation templates toward further constraining the rate of occultations by kilometer-scale KBOs and thus their overall abundance.

\section{Observations}

The design of a serendipitous occultation survey generally aims to monitor as many stars for as long as possible within available resource limits to maximize sensitivity. KBOs are additionally concentrated around a mean plane tilted ${\sim}1^\circ\llap{.}7$ from the ecliptic plane \citep{brown2004}, so the KBO occultation rate will be higher for stars near this plane. Additionally, KBOs are at a distance ($\sim$40~au) where effects from both diffraction and the apparent diameters of occulted stars become pronounced for kilometer-scale occulters \citep{roques1987,roques2006}, requiring the use of \edit{2}{bright stars with high signal-to-noise (S/N) ratios and angularly small disks, in order} to detect occultations by such small and distant objects. In exchange, both size (largely from occultation depth) and distance (largely from occultation duration) can be simultaneously extracted from a single light curve with sufficiently high S/N, whereas nearly geometric occultations by nearer objects constrain only the size.

Balancing these factors, we selected a dense star field centered on the \object{M22} globular cluster at a distant $\sim$3~kpc \citep{gaia2023} away just $0^\circ\llap{.}7$ south of the ecliptic and $0^\circ\llap{.}3$ south of the Kuiper Belt mean plane measured by \citet{matheson2023}. While the predominantly old stars tend to be cooler and thus angularly large for their brightness, only a few dozen of the largest stars are angularly larger \citep{gaia2023} than the optical Fresnel scale \citep{roques2000} of $\sim$1~km $\sim0.05$~mas at $\sim$40~au away. The smearing effect of nonzero stellar diameter on the occultation is therefore dwarfed by the diffraction effects for the overwhelming majority of stars in the cluster.

\begin{figure*}
\includegraphics[width=\linewidth]{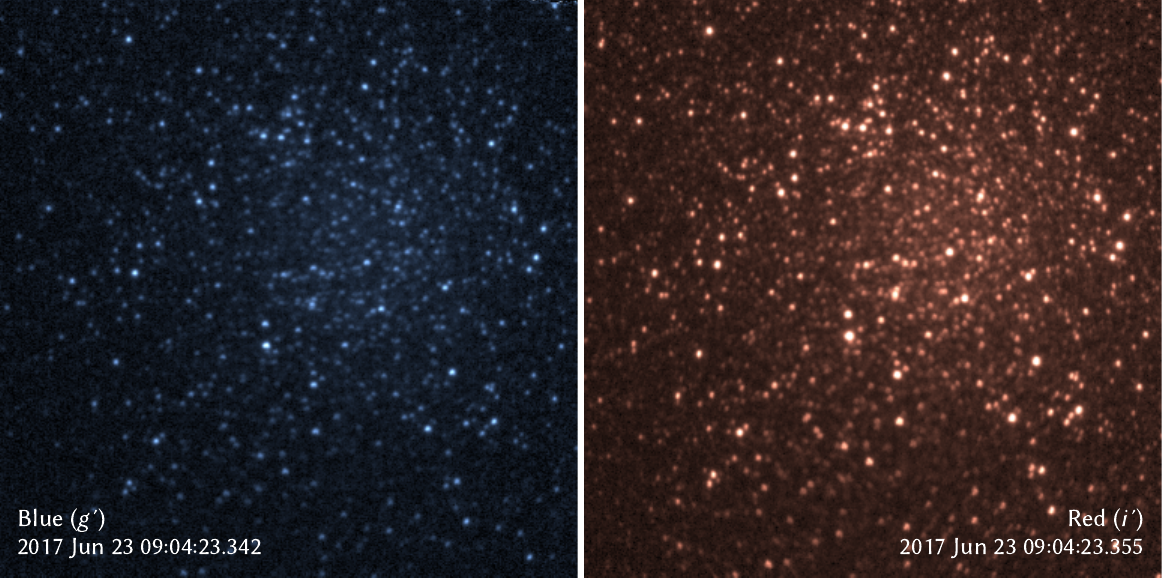}
\caption{Example of nearly simultaneous 30.39~ms exposure frames of M22 through the blue ($g'$; left) and red ($i'$; right) channels after bias and flat calibration. The $g'$/$i'$ images in this and subsequent figures are tinted blue/red for clarity.}
\label{fig:m22}
\end{figure*}

CHIMERA's $5'\times5'$ field-of-view covers only the core of the cluster but already captures several thousand distinguishable stars in single 30.39~ms exposures through both the blue/$g'$ and red/$i'$ channels, as demonstrated in Figure~\ref{fig:m22}. We ran CHIMERA with $2\times2$ binning at a plate scale of $0''\llap{.}6$~px$^{-1}$, which enabled a frame rate of $\sim$33~Hz (30.39~ms exposure time + 0.67~ms overhead between frames), and successfully observed the M22 field for a total of 63.1~hr spread over 24 nights, as summarized in Table~\ref{tab:obs}.

\begin{deluxetable}{lcc}
\tablecaption{Summary of Observations}
\label{tab:obs}

\tablecolumns{6}
\tablehead{
\colhead{Date} & \colhead{M22 Data} & \colhead{Seeing} \\
\colhead{(UT)} & \colhead{(hr)\tablenotemark{a}} & \colhead{(arcsec)\tablenotemark{b}}
}

\startdata
2015 Jul 12 & 3.4 & 1.0\\
2015 Jul 13 & 3.5 & 1.1\\
2015 Jul 14 & 3.3 & 1.2\\
2015 Jul 15 & 3.5 & 1.1\\
2015 Jul 16 & 4.1 & 1.0\\
2015 Aug 14 & 3.0 & 1.1\\
2015 Aug 15 & 1.7\tablenotemark{c} & 1.4\\
2015 Aug 16 & 3.1 & 1.4\\
2015 Aug 17 & 2.4 & 1.2\\
2016 Jul 4 & 3.3 & 1.5\\
2016 Jul 28 & 0.6\tablenotemark{c} & 1.3\\
2016 Jul 29 & 2.8 & 1.4\\
2016 Jul 30 & 2.7 & 1.1\\
2016 Jul 31 & 2.2 & 1.0\\
2016 Aug 1 & 3.3 & 1.3\\
2016 Aug 2 & 2.8 & 0.8\\
2016 Aug 4 & 1.0\tablenotemark{c} & 1.2\\
2017 Jun 23 & 3.3 & 1.8\\
2017 Jun 24 & 1.2\tablenotemark{c} & 1.8\\
2017 Jun 25 & 2.7 & 1.8\\
2017 Jun 26 & 1.2 & 1.7\\
2017 Jul 21 & 2.3 & 1.2\\
2017 Jul 22 & 3.0 & 1.5\\
2017 Jul 23 & 2.5 & 1.1\\
Total & 63.1 & \nodata
\enddata

\tablenotetext{a}{Hours of integration on M22 per night in each of $i'$ and $g'$.}
\tablenotetext{b}{Nightly average zenith seeing FWHM reported by seeing monitor.}
\tablenotetext{c}{Partially disrupted by clouds.}
\end{deluxetable}

Observations were conducted similarly across all nights, with the telescope pointed at M22 at the start of the sequence and left to sidereally track without closed loop guiding for the duration of the observation period, over which the pointing typically drifts by on the order of ${\sim}1'$. Also, most observations were collected at an airmass of ${\sim}2$ as limited by the southern declination of M22 and the northern latitude of Palomar Observatory.

\section{Data Reduction}
\label{sec:data}

The process of transforming the image cubes of M22 into KBO occultation constraints is divided into two major steps: (1) extracting the light curves of the stars from each frame with a technique specifically adapted to the extremely dense star field of M22's core, and (2) a general approach of matching the processed light curves against templates of light curve patterns expected for true occultations by KBOs to place limits on the occultation rate, and subsequently the KBO number density. These steps are detailed in sections~\ref{subsec:phot} and \ref{subsec:occ}, respectively. For our present analysis, we used only the more sensitive red/$i'$ channel data to set occultation rate and thus KBO number density constraints, and reserved the blue/$g'$ data to subsequently evaluate the consistency of the limiting events with the wavelength-dependent light curve patterns expected for true occultations.

\subsection{Photometry}
\label{subsec:phot}

The survey data set presents a couple of unique challenges to be overcome for accurate photometry:

\begin{enumerate}
\item The M22 star field packs many thousands of stars into a single $5'\times5'$ CHIMERA field of view. While most are indistinguishable from being blended with brighter, adjacent stars or otherwise too faint for CHIMERA to detect KBO occultations of, their collective presence precludes the use of classical aperture photometry methods as any circular aperture or annulus will be contaminated by a numerous neighboring stars.
\item The full data set contains several million individual frames, each of which requires the photometric extraction of several thousand stars. Traditional dense-field photometry methods involving the direct fitting of point spread functions (PSFs) are computationally intensive and cannot be practically applied to the full data set with our available computing resources.
\end{enumerate}

We developed a custom photometry method that combines both the classical aperture photometry and PSF-fitting techniques to reliably and efficiently extract dense-field photometry from a fixed field-of-view.

Our data were written in cubes of 1000 frames ($\sim$30~s; much longer than the $<$1~s duration of targeted occultations) each, which we processed largely separately. However, we first selected one of these image cubes from the middle of the observation sequence of each night to serve as the reference cube for that night. We applied standard bias frame and flat field corrections to all frames of this reference cube, corrected for the slight ($\sim$1\%) gain instability in recorded frames by normalizing to the median of all pixels in each frame, then used cross correlation minimization to align and shift all frames to correct telescope tracking and atmospheric tip-tilt effects. We then repeated the process for all image cubes from that night, aligning and shifting them to the pointing of the reference field such that all frames collected on each night are co-aligned.

Next, we astrometrically solved for the center and orientation of the reference field using a third-order polynomial distortion model previously derived by \citet{harding2016}. We then used the astrometric solution to select the 15 brightest stars contained in each field that are isolated $>$5~px ($3''$) horizontally and vertically from any other stars less than 2~mag fainter in order to generate an $11\times11$~px effective point spread function \citep[ePSF,][]{anderson2000} from the mean stack of each image cube, as well as from each frame individually. From the mean ePSF, we generated a simulated frame with astrometry matching the mean, containing all stars from the catalog within the observed field-of-view, and scaled the simulated star frame jointly with a flat sky background plus a 2D Gaussian distribution representing unresolved stars excluded from the catalog to photometrically fit the mean observed frame.

For astrometry, photometry, the generation of simulated frames, as well as star identification numbering, we used the M22 star catalog by \citet{liu2015}, which provides $\sim$12,000 total stars within our field-of-view---several times more than the Gaia DR3 catalog, which has poor completeness due to the high star density in the cluster \citep{gaia2023}. This catalog provides Johnson $V$ magnitudes and Str\"{o}mgren $b-y$ colors, which we converted to SDSS $g'$ and $i'$ magnitudes using the transformations described in \citet{cousins1987} and \citet{jordi2006}. These transformed magnitudes, however, can have sizable errors due to a combination of stellar properties not captured in the simplistic transformation functions and actual, long-term physical variability in the stars themselves.

Therefore, for each night, we performed a first iteration of light curve extraction on the reference cube alone to measure the mean stellar fluxes through our actual filter bandpasses. For each star in the catalog within the field, we performed the following steps:

\begin{enumerate}
\item We determined an optimal aperture for the star using the initial simulated frame and the signal contribution of the star by starting the nearest pixel to the position of the star and its eight surrounding pixels, then iteratively testing the expected S/N of all adjacent pixels, adding only those pixels that increase the expected S/N into the aperture and repeating until no additional pixels can be added to the aperture to increase the expected S/N.
\item We used the initial simulated frame to estimate the total background flux from sky brightness plus all other resolved and unresolved stars contained within the aperture, taking the flux of the star to be the flux contained within the aperture minus the estimated total background flux.
\end{enumerate}

\begin{figure*}
\includegraphics[width=\linewidth]{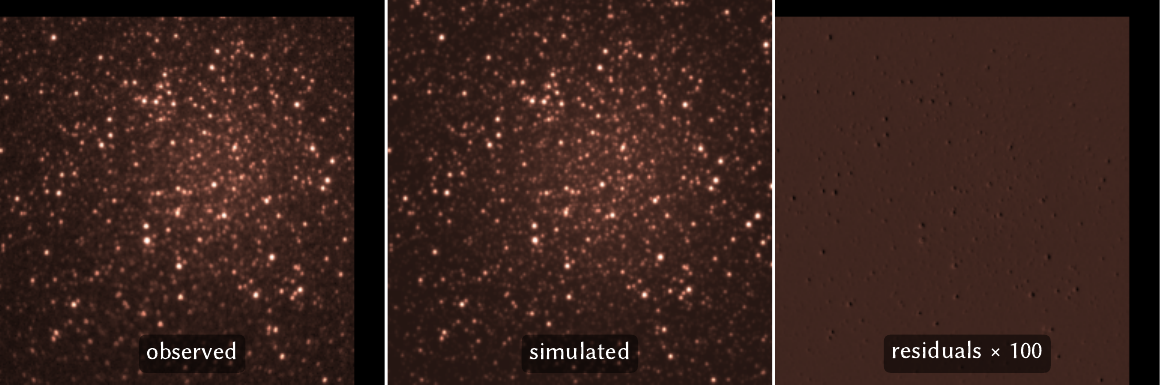}
\caption{Demonstration of simulated frame matching, showing the red channel ($i'$) frame in Figure~\ref{fig:m22} (left), the simulated model frame (center), and the observed minus simulated residual frame (right), with the residuals (on the order of only $\sim$0.1\% of the peak flux) amplified by $100\times$ for visibility. Note that the observed frame has been shifted into alignment with the simulated frame centered on the field of the nightly reference frame, from which the true camera pointing slightly drifts over the course of each night.}
\label{fig:sim}
\end{figure*}

This procedure produces a set of optimal apertures and mean fluxes for all catalogued stars on the frame. We then considered only the subset of stars whose light curves have S/N $>5$ (as measured by the mean flux divided by the standard deviation), corrected the transformed catalog magnitudes of these stars to match their measured mean fluxes, then repeated the photometry process for each of these stars in all image cubes with a new set of simulated frames using the ePSFs of every individual frame. Figure~\ref{fig:sim} illustrates the closeness of the match between an observed frame and its corresponding simulated frame.

The match, however, is still not perfect with clear systematic errors that vary on a frame-to-frame basis, including variations in the true PSF across the frame, as well as temporal variation in the image distortion from atmospheric effects. These errors then translate into artifacts in the measured light curve, which can sometimes appear similar to patterns expected from true occultations. Such events, however, can often be distinguished from occultations by their tendency \edit{1}{to} simultaneously affect many stars at once, e.g., when a particularly turbulent patch of atmosphere crosses the field, so can in principle be identified by searching for times when a large fraction of light curves all significantly stray from their mean values.

To quantify this source of uncertainty, we \edit{1}{took} all the light curves from each image cube and first \edit{1}{normalized} them to a mean of zero and a standard deviation of one, i.e., into a times series of $z$-scores. \edit{1}{Next, we compared these light curve $z$-scores across all stars at every individual frame/time. When a patch of turbulence (or a similar effect) affects at least a portion of the frame, the ensemble of light curve $z$-scores from the corresponding frames/times exhibits a much larger spread than the standard normal distribution (since these $z$-scores of focus are ones normalized to the standard deviations of the individual light curves, and not to that of the ensemble of values across light curves). We originally used the standard deviation of these $z$-scores directly as the standard error at each frame/time in the normalized light curves. However, we found that in practice, the light curve $z$-score ensembles often deviate substantially from normal distribution during periods of high turbulence/similar events, and the ordinary standard deviation fails to adequately capture the much larger excursions exhibited a small number of stars/light curves than expected by the normal distribution. These photometrically poor \edit{2}{time} periods appear to cover only a small fraction of the total observing time, so we chose to simply minimize \edit{2}{the contributions of these periods of all light curves at each frame/time to reflect the spread of all normalized light curves at that frame/time.}}

\edit{1}{More specifically, we used the stars with the second highest and lowest $z$-scores across the ensemble of all light curves at each frame/time. We chose to consider the second highest/lowest values, so the procedure does not downweight \edit{2}{a light curve} with \edit{2}{a} deep, real occultation, if one is responsible for the highest/lowest $z$-score. It would be unlikely for multiple such occultations to occur simultaneously, and consider the second highest/lowest $z$-scores to reflect only noise. These raw values, however, would not be appropriate to use as the standard errors, because the second highest/lowest values of a large sample drawn from even a normal distribution will be several $\sigma$ from the mean, e.g., the second highest/lowest values of a normally distributed sample of 1000 items will fall near $\pm3\sigma$. We then considered the normal distributions whose quantile functions match the second highest/lowest $z$-scores stars at their respective quantiles (i.e., $1-1.5/N_\star$ and $1.5/N_\star$ for the second highest and lowest values out of $N_\star$ light curves, respectively), and used the mean of the standard deviations of these two matched normal distributions as the standard error for the frame/time. Finally, we applied a shift to the nominal flux at each frame/time to set the mean $z$-score of all light curves at each frame/time to zero, as a final step to reduce the degree of artifacts correlated between multiple light curves.}

\begin{figure}
\includegraphics[width=\linewidth]{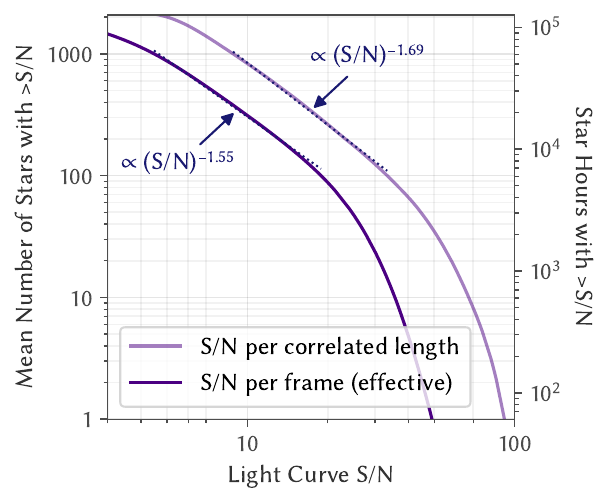}
\caption{Cumulative distribution of stars by effective S/N measured per correlated length of light curve (light blue) and per 30.39~ms exposure frame (dark blue). The distributions roughly follow a truncated power law, with the mean number of stars per frame above a given S/N scaling approximately with $(\text{S/N})^{-1.6}$ at between 100 and 1000 stars.}
\label{fig:snr}
\end{figure}

The resulting, processed light curves still exhibit a great degree of correlated noise, which we have quantified by fitting an exponential decay model to the autocorrelation of every light curve. Across our full data set, light curves exhibit an average, effective correlation length of $\sim$4 frames ($\sim$0.1~s), corresponding to a factor of $\sim$2 loss in S/N in time-binned light curves compared to a similar light curve with fully independent (e.g., readout or shot noise limited) points. Figure~\ref{fig:snr} illustrates the distribution of both the raw light curve S/N and the equivalent, correlation-corrected per-frame S/N within our M22 data set, showing that the number of stars in each frame with a light curve above a particular S/N roughly follows a power law $\propto(\text{S/N})^{-1.6}$ in the range of 100--1000 stars.

\subsection{Occultation Search}
\label{subsec:occ}

\subsubsection{Occultation Templates}

Due to their size and distance, occultations by kilometer-scale KBOs are significantly altered by diffraction effects, so are poorly represented by a near-instantaneous total disappearance and re-emergence of the occulted star \citep{roques1987}. Instead, diffraction effectively sets a minimum radius for the KBOs' shadows that is approximately the Fresnel scale of $\sim$1~km \citep{roques2000}, with the physical sizes of smaller KBOs affecting only the depth rather than the size of the shadow, and also produces a ringing pattern of both positive and negative brightness excursions extending well beyond the shadow core. We used the analytic model of \citet{roques2000} for the occultation shadow of KBOs modeled as circular disks, \edit{1}{and accounted for smearing of the shadow from nonzero stellar diameters by convolving the analytic model with stellar disks approximated as circular and uniformly bright}.

The light curve of an occultation is a 1D cross section through this 2D diffraction shadow, scaled for the relative speed between the occulter and observer. To generate this cross section \edit{1}{of the wavelength-dependent shadow}, we \edit{1}{performed} Monte Carlo sampling of wavelengths over 695--844~nm for $i'$ and 401--550~nm for $g'$, approximated with uniform weighting over those intervals, i.e., neglecting the higher order impact of star color. Our observations of M22 were predominantly made near opposition, where the relative speed of KBOs is predominantly set by Earth's 30~km~s$^{-1}$ orbital speed around the Sun, which dwarfs the $\sim$5~km~s$^{-1}$ circular orbital speed of KBOs at $r\sim40$~au from the Sun. Nonetheless, given the simplicity of the correction, we modeled the relative speed for the actual viewing geometry at each epoch, with KBOs treated as moving parallel to the ecliptic in circular, heliocentric orbits.

From this model, we can simulate the light curve signature of occultations with any given KBO heliocentric distance $r$, diameter $D$, occultation impact parameter (minimum linear distance of the observer from the shadow center) $b$, and stellar angular diameter $\delta$. The $\delta$ is fixed for each star, which we computed by estimating its effective temperature from its $g'-i'$ color using \citet{fukugita2011}, then crudely approximating its emission as blackbody and treating all stars as being at common distance of 3~kpc, with the understanding that errors in $\delta$ of several tens of percent, or even a factor of a few, only minimally affect the actual diffraction shadow as the stellar disk is dwarfed by the angular Fresnel scale for the vast majority of observed stars. The other three parameters---$r$, $D$, and $b$---can take on a wide range of values that will vary between events.

For our survey of occultations by kilometer-scale KBOs, we targeted a cube of parameter space encompassing at least $35~\text{au}<r<50~\text{au}$ \citep[covering most classical KBOs, e.g.,][]{bannister2018}, $0.5~\text{km}<D<2~\text{km}$, and $b<1.5~\text{km}$, and set up a grid of occultation templates covering the region against which the observed light curves can be compared. We experimented with various grid densities, evaluating each choice comparing the root-mean-squared (RMS) \edit{1}{differences} of all adjacent templates on the grid\edit{1}{---calculated by subtracting one template from another, squaring the resulting points, and taking the square root of the mean of the result (i.e., the RMS residuals of using templates to fit adjacent templates acting as simulated occultations)---and that of the template with an occultation-free flat line}. We settled on the following grid points:

\begin{enumerate}
\item $r=$~35~au, 40~au, 45~au, 50~au
\item $D=$~0.5~km, 0.6~km, 0.7~km, 0.8~km, 1.0~km, 1.2~km, 1.4~km, 1.7~km, 2.0~km
\item $b=$~0~km, 0.5~km, 1.0~km, 1.5~km
\end{enumerate}

With this grid, $>$90\% of all pairs of adjacent templates have RMS \edit{1}{differences} $<$10\% that of the RMS \edit{1}{difference} of either compared with the occultation-free flat line, in $i'$. \edit{1}{An exact match of an occultation with a template of its actual properties provides the maximum S/N, and corresponds to a template compared to itself, with zero RMS difference. Matching any occultation with a flat line yields a zero S/N match. Both signal (and thus S/N) and RMS difference scale linearly with occultation/template amplitude, so this set of templates generally provides} $<$10\% loss of S/N fitting an occultation on a grid point with the neighboring template. We treat each of these templates as representing the centers of their respective bins fully covering the entire spanned region of parameter space, with each bin separated from the next at the midpoint of the respective templates' parameters and endpoint bins treated as symmetric about their central value (e.g., the $D=2.0$~km bins cover 1.85--2.15~km), aside from $b=0$~km which is treated as covering 0--0.25~km since $b\geq0$ by definition. We computed all occultation templates out to a radius of $3\times$ the largest Fresnel scale of occultations on the grid, set by the $r=50$~au templates, which typically corresponds to an occultation duration of 18~frames $\sim0.5$~s in $i'$.

\subsubsection{Template Fitting}

Demonstrating that an occultation-like signature in an observed light curve actually arises from an occultation rather than atmospheric turbulence artifacts amplified by dense-field photometric errors is a difficult challenge requiring a detailed model accurately capturing not only how those effects typically behave, but also their extreme, low probability tail behavior at the frequency of the targeted occultation rate. Such a model could be obtained, for example, with a dedicated, comparably-sized control data set targeting stars angularly too large to be occulted by kilometer-sized KBOs, or near the ecliptic poles where few KBOs are present. The results of such an undertaking, however, only become useful if the main survey has identified plausible candidate occultations that cannot be ruled out as occultations by other means.

In fact, demonstrating that a light curve does \emph{not} contain a particular occultation signature is a far simpler task that requires only a straightforward statistical test comparing the observed light curve with a template of that occultation signature, with the null hypothesis that the observed light curve arises from that template plus noise. The corresponding $p$-value is the conditional probability for that section of light curve to be at least as different from the template as it is given the assumption that an occultation matching the template did actually occur there. Any given real occultation is unlikely to be disrupted by \edit{1}{occasional, high-amplitude fluctuations arising from atmospheric or related systematic effects}, so we can perform this test using only the typical noise characteristics of the light curve, quantifying the distance between the light curve and template by the Mahalanobis distance \edit{1}{($\mathbf{d}^{-1}\mathbf{C}\mathbf{d}$ for residuals vector $\mathbf{d}$ and covariance matrix $\mathbf{C}$)} using the flux uncertainties and autocorrelation measurements previously described in section~\ref{subsec:phot}. We then estimate the expected distribution of this template--light curve distance for real occultations by injecting the template everywhere along the light curve and measuring the resulting distribution of distances.

\begin{figure*}
\includegraphics[width=\linewidth]{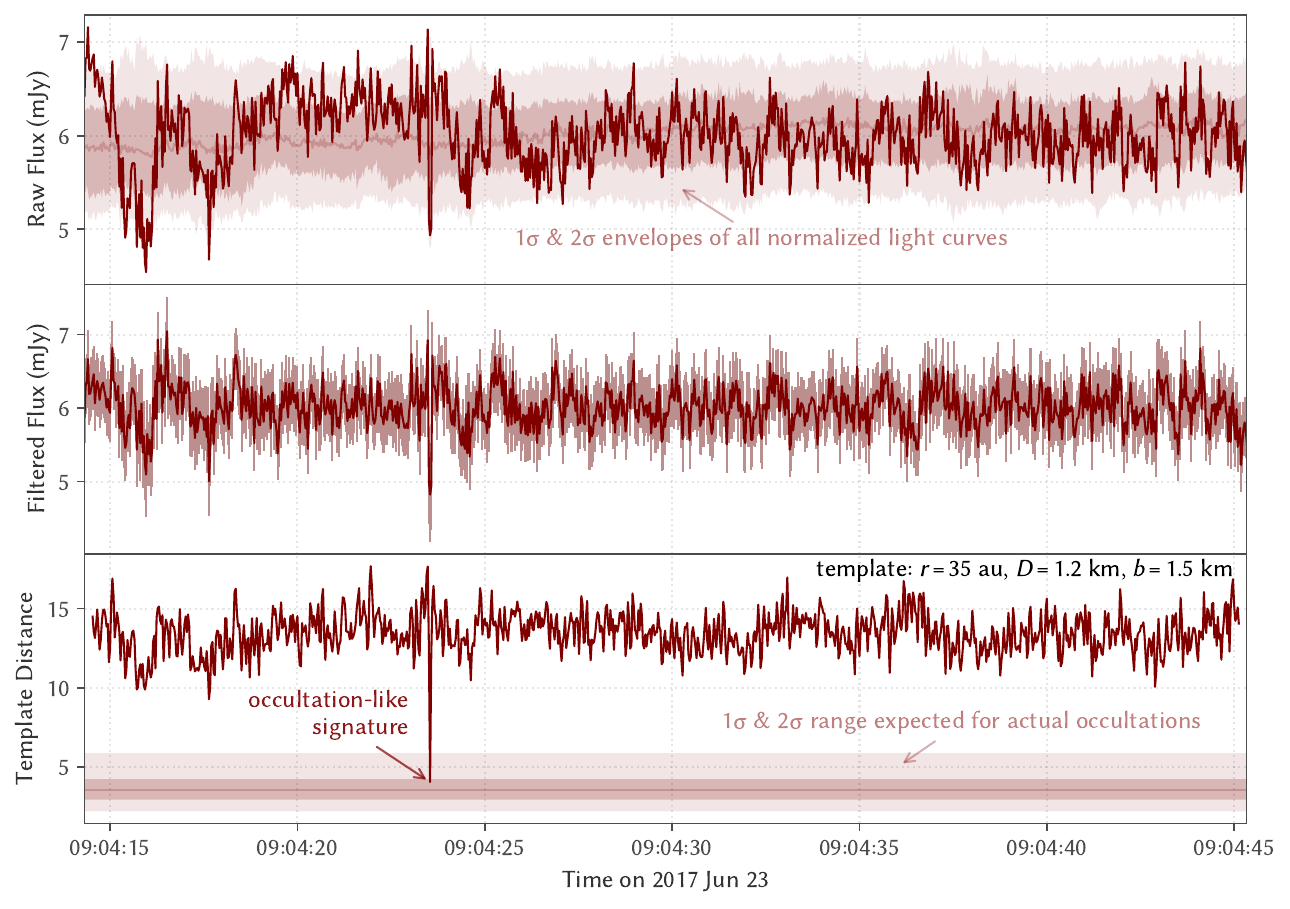}
\caption{Example of an $i'$ light curve being fit with a template. Top: The raw light curve for star 2636 compared with the envelope of light curves of other stars in the field normalized to the same brightness and S/N. Middle: The same light curve passed through a high-pass filter, corrected with the mean offset of the other normalized light curves with error bars set by the spread of those light curves. Bottom: The goodness of fit of an occultation template sliding along the light curve, as measured by the Mahalanobis distance between the observed light curve and template at each point based on the autocorrelation measured from the light curve. The $1\sigma$ and $2\sigma$ ranges of the template distance expected from simulated, injected occultation events are indicated here. An inset of the labeled occultation-like signature in this light curve alongside the associated template is shown in the upper row of Figure~\ref{fig:lev}.}
\label{fig:lc}
\end{figure*}

Figure~\ref{fig:lc} illustrates the full process, starting with the raw light curve of a star from an image cube with uncertainties derived from the light curves of other stars, then passing it through a high-pass filter to remove features \edit{1}{longer than the} occultation template, and finally sliding the template along the processed curve \edit{1}{to compute the template--light curve distance at each point along the light curve. This process quantifies the similarity of the light curve to a template, identifying locations where that similarity (i.e., the template--light curve distance) is comparable to values expected if there were a real occultation with the template's parameters at that position. Real occultations will likewise not match the template exactly due to the presence of noise, so will have a nonzero template--light curve distance distributed by a $\chi^2$-like probability distribution. We quantify this expected distribution by adding (``injecting'') the template at each point along the light curve to represent the appearance of a true occultation. We then take the resulting collection of template--light curve (with injected occultation) distances as the probability distribution for the value of the template--light curve distance a real occultation with the parameters of the template would take, if such an occultation were actually present on the light curve. Under this formalism, light curves farther from/more dissimilar to the template than all except a false negative fraction $\beta$ of injected occultations can be ruled out as containing a real occultation to a confidence level of $1-\beta$.}

\subsubsection{Constraining the Occultation Rate}

We use the template--light curve distance to define another S/N: that of the light curve with respect to each template. Here, we consider the ``signal'' to \edit{1}{be} the difference in the mean template--light curve distance of the actual light curve (i.e., the expected value of the template--light curve distance in the absence of an occultation) and that with an occultation injected (i.e., the expected value of the template--light curve distance if an occultation matching the template were present). Meanwhile, we consider the ``noise'' of this template S/N to be the quadrature sum of the standard deviation of the template--light curve distance \edit{1}{along the actual light curve (i.e., the data curve in the bottom panel of Figure~\ref{fig:lc}), and that of the light curve with injected occultations (i.e., the spread of the probability distribution of the template--light curve distance of a real occultation, corresponding to the indicated range expected for actual occultations in the bottom panel of Figure~\ref{fig:lc})}. Then, for each template or set of templates spanning the range of occultation parameters of interest, we constrain the overall occultation rate by sorting all the light curves by their template S/N, counting the number of star hours of light curves from which occultations can be rejected starting from highest to lowest S/N until reaching a light curve our selected threshold is unable to reject as containing an occultation, and assuming the occultations follow Poisson statistics, with the final rate corrected for the permitted $\beta$.

Under this method, the choice of a $\beta$ threshold significantly affects the overall survey sensitivity. Enforcing a very low $\beta$ ensures with high confidence that light curves with no template--light curve distances meeting the threshold do not contain any occultations with the template's parameters, but is only capable of ruling out occultations from high S/N light curves, limiting search depth. In contrast, permitting a higher $\beta$ will expand the number of usefully searchable light curves at the cost of allowing that fraction of real occultations to slip through, which requires a larger \edit{1}{correction factor $(1-\beta)^{-1}$} to the occultation rate limit that likewise harms sensitivity.

\begin{figure}
\includegraphics[width=\linewidth]{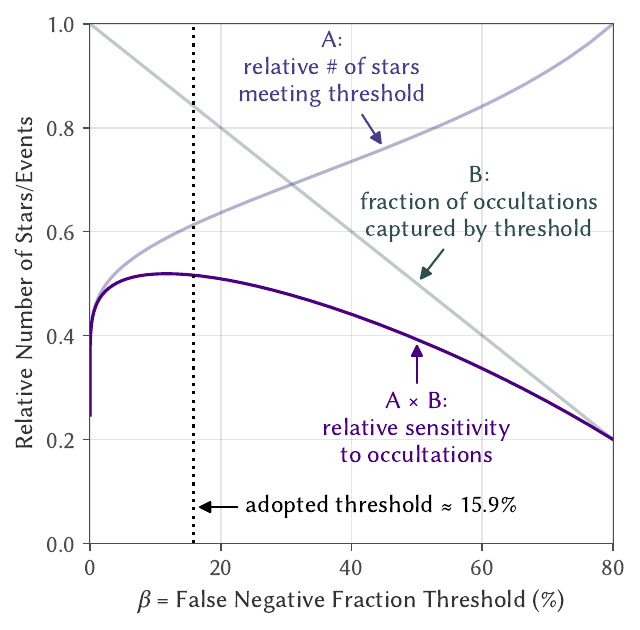}
\caption{Comparison of the estimated number of included stars vs. fraction of included occultations for a range of false negative fraction thresholds $\beta$, showing an optimum in relative sensitivity (the product of the two quantities) near the chosen $\sim$15.9\% (i.e., $1\sigma$ level of the normal distribution).}
\label{fig:fnf}
\end{figure}

Figure~\ref{fig:fnf} illustrates the balance between these two factors in selecting a threshold that allows a large number of stars/light curves to be used without losing \edit{2}{too many of the true occultations present within those light-curves. In other words, we aim to maximize} the effective number of stars of usable stars hours of light curve---the number of star hours of light curves with a high S/N beyond the threshold contributing to the constraint, corrected for the \edit{1}{chosen} $\beta$. Assuming the number of stars above a given S/N scales as $(\text{S/N})^{-1.6}$, as previously estimated in section~\ref{subsec:phot}, we estimate an optimal sensitivity near $\beta\sim5$--20\%, with only a few percent difference over this range. We opted for a value of $\sim$15.9\% at the $+1\sigma$ limit of template--light curve distance for true occultations. With this selection, any light curve in which an occultation cannot be rejected by this relative weak threshold should qualitatively appear to be well-fit to the template, and the limiting light curves setting the occultation rate limits conveniently become the well-fit occultation candidates with the highest S/N for further evaluation in section~\ref{sec:results}.

\subsubsection{Occultation Rate to KBO Sky Density}

In order to convert the constraint on occultation rate into one on the actual sky density of KBOs, we consider the scanned area of sky corresponding to the high S/N light curves in which occultations can be rejected, which is computed from the relative speed $v_i$ of a KBO at geocentric distance $d_i$ (approximately equal to heliocentric distance $r_i$), a template impact parameter bin width $\Delta b_i$, and duration $\Delta t_i$ of the light curve segment with the $i^{\text{th}}$ highest template S/N satisfying the $\beta$ threshold as

\begin{equation}
\Omega=\sum_i\frac{v_i\times\Delta b_i\times\Delta t_i}{d_i}
\end{equation}

\noindent where we have treated scans of the same star/light curve by templates of different $b$ equivalently to separate stars, since each additional range of $b$ for which an occultation can be rejected also rules out the presence of a corresponding KBO in an additional parallel, band of sky.

If the limiting event---the highest template S/N light curve segment in which an occultation cannot be rejected---is considered a true occultation, then the sky density $\Lambda$ of KBOs with the parameters of the searched templates is nominally $\langle\Lambda\rangle=\Omega^{-1}$, since the area of sky searched before encountering the first occultation follows a probability distribution $\Lambda\exp(-\Lambda\Omega)$ under this Poisson process. Likewise, the $\Lambda$ that places $\Omega^{-1}$ at the $95^{\text{th}}$ percentile of this probability distribution, equal to the 95\% confidence upper limit on the true $\Lambda$, is $\Lambda_{95}\approx3\Omega^{-1}$, which we adopt as a robust constraint irrespective of whether the corresponding limiting event is a true occultation.

\section{Survey Results}
\label{sec:results}

\subsection{Constraints on KBO Sky Density}

\begin{figure}
\includegraphics[width=\linewidth]{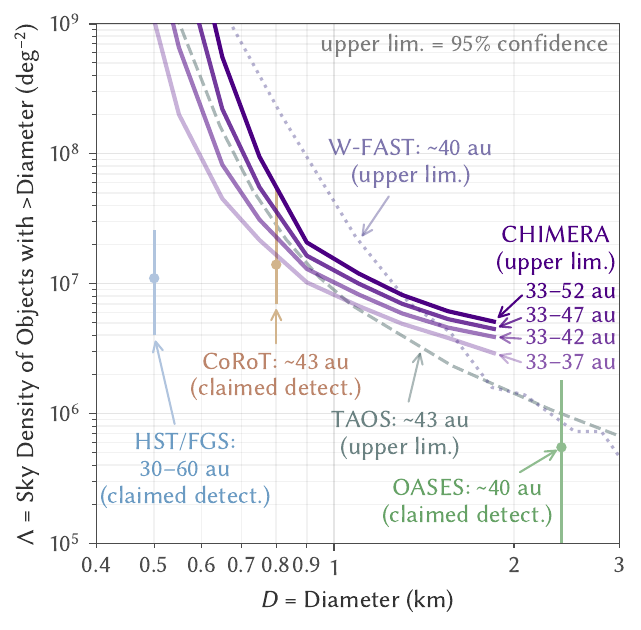}
\caption{The 95\% confidence upper limits for the ecliptic sky density of kilometer-scale KBOs larger than the plotted diameters, compared with earlier limits set by TAOS \citep{zhang2013} and W-FAST \citep{nir2023}, as well as the claimed detections from HST/FGS \citep{schlichting2012}, CoRoT \citep{liu2015}, and OASES \citep{arimatsu2019}. The labeled distances indicate the approximate range of KBO distances covered by each survey/upper limit curve.}
\label{fig:lim}
\end{figure}

We computed the cumulative 95\% confidence upper limits on sky density for all KBOs larger than various $D$ within various $r$, and plotted the resulting set of upper limit curves in Figure~\ref{fig:lim}. Note that we have little sensitivity to occultations by KBOs larger than the $D\sim1.8$--2.2~km covered by the 2~km template bin, the contribution of these larger KBOs to the cumulative sky densities is negligible if the size distribution follows a power law $N({>}D)\propto D^{-s}\equiv D^{1-q}$ (i.e., cumulative power index $s$ and differential index $q$) extending to the sky density of KBOs observed by direct telescopic surveys. We corrected our limits assuming a break at $D=90$~km with $\Lambda({>}90~\text{km})=5.4$~deg$^{-2}$ as measured by \citet{fraser2008}, but found that any $\Lambda({>}90~\text{km})\lesssim10^6$~km$^{-2}$ negligibly affects the results, so our upper limits remain effectively independent of any realistic size distribution model.

Our derived 95\% confidence upper limit of $\Lambda({>}1~\text{km})<\Lambda_{95}({>}1~\text{km})\sim10^7$~deg$^{-2}$ is comparable to the corresponding upper limits set by TAOS \citep{zhang2013} and W-FAST \citep{nir2023}, as well as the density implied by the claimed detections by CoRoT \citep{liu2015}. We did not search for occultations as large as the claimed detection by OASES \citep{arimatsu2019}, although our power law extension to $\Lambda({>}90~\text{km})=5.4$~deg$^{-2}$ of \citet{fraser2008} yields a 95\% upper limit on the differential index of $q\equiv s+1<4.3$ for the full range of $r=33$--52~au spanned by our template bins. The claimed detections by HST/FGS \citep{schlichting2012} primarily constrain the sky density of KBOs around $D=0.5$~km which create occultations with insufficient S/N to be detected for most of our light curves, so we do not have a comparable constraint at that size.

We caution, however, that although we have placed all of these constraints together on the same plot for reference, the precise numbers may not be directly comparable. In addition to being derived through different methods with different standards for interpretation of results, the various surveys also differ in the definition of sky density being measured, with most targeting KBOs around $r=40$~au or 43~au, often without specifying the actual sensitivity range. Occultations appear fairly similar over the relatively narrow range of considered $r$, although our results do show slightly weaker constraints when searching for occultations over a wider range out to larger $r$ from a couple of effects:

\begin{enumerate}
\item Searching a larger region of parameter space provides more opportunity for noise to create occultation-like signatures. For example\edit{2}{,} a noisy light curve might coincidentally match an $r=45$~au template sufficiently well for the corresponding occultation to not be rejected while differing sufficiently from all other templates for those occultations to be rejected, which weakens the 33--47~au and 33--52~au upper limits without affecting the 33--37~au and 33--42~au limits.
\item \edit{2}{More} distant KBOs create shallower, longer duration occultations that more easily blend in with the strongly correlated, atmospheric noise, so \edit{2}{they} require light curves to have higher S/N to reject than do occultations by nearer KBOs.
\end{enumerate}

The surveys also target stars over different ranges of ecliptic/Kuiper Belt mean plane latitude, where the actual sky density of KBOs may vary. Our M22 dataset contains only light curves from stars only $0^\circ\llap{.}3$ from the Kuiper Belt mean plane. OASES \citep{arimatsu2019} similarly monitored a single field of stars within $2^\circ$ of the ecliptic. TAOS \citep{zhang2013} used a sample of stars 90\% within $6^\circ$ of the ecliptic, while \edit{1}{W-FAST} \citep{nir2023} derived their limit from a sample of stars with 95\% within $4^\circ$ of the ecliptic. The CoRoT \citep{liu2015} and HST/FGS \citep{schlichting2012} surveys used stars spread as far as several tens of degrees from the ecliptic, and corrected their constraints to ecliptic sky densities based on modeled inclination distributions of kilometer-scale KBOs. Given the uncertain model-dependence of such corrections and differences in methodology across the studies, we have not attempted to correct any of these other reported constraints to match our results, which we instead simply consider to provide a related but separate upper limit.

\subsection{Limiting Events}

\begin{figure*}
\centering
\raisebox{-\height}{\includegraphics[width=0.4\linewidth]{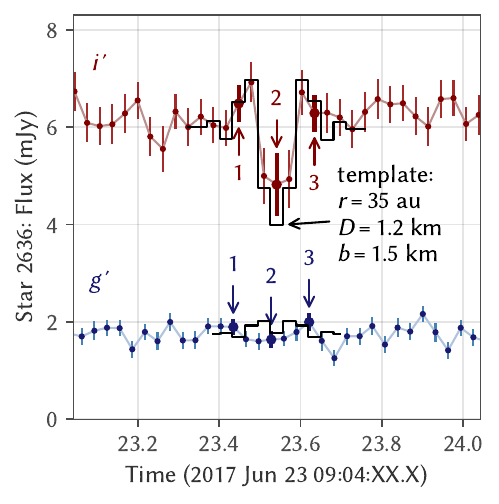}}
\raisebox{-\height}{\includegraphics[width=0.59\linewidth]{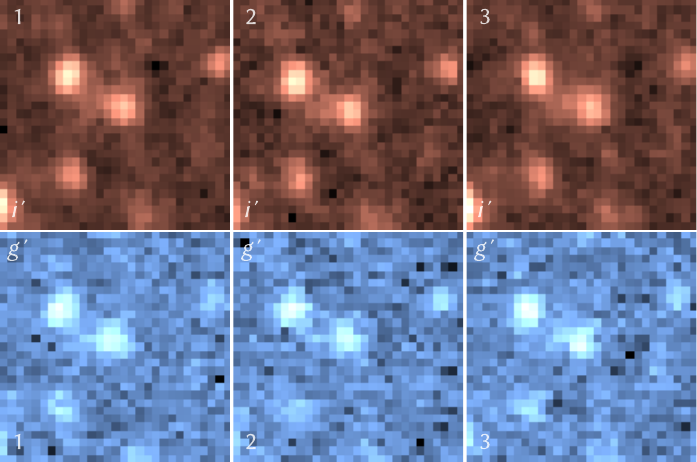}}
\raisebox{-\height}{\includegraphics[width=0.4\linewidth]{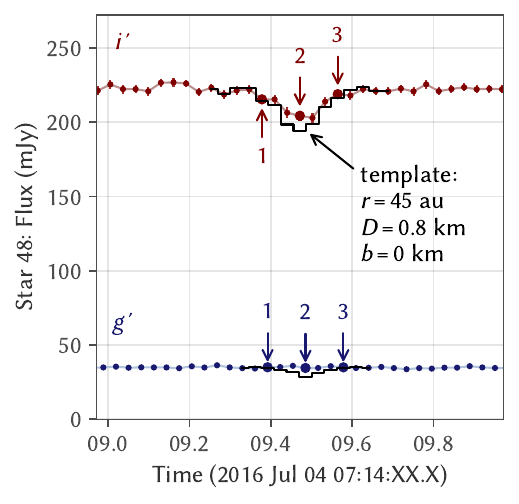}}
\raisebox{-\height}{\includegraphics[width=0.59\linewidth]{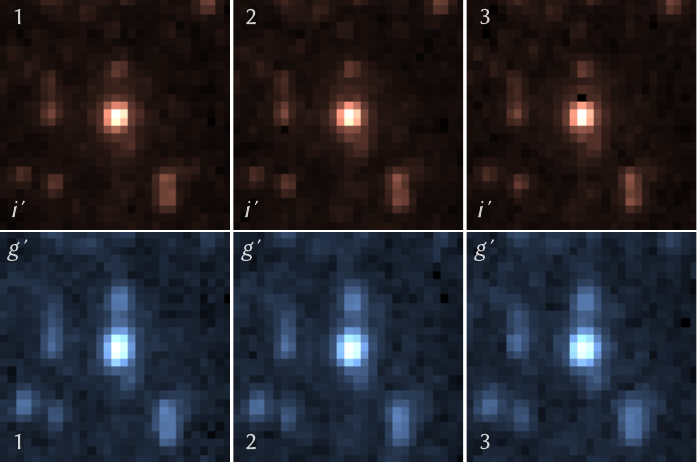}}
\caption{\edit{1}{Two examples of limiting events showing occultation-like signatures in $i'$ light curves,} setting the upper limits on \edit{1}{the total 33--52~au} KBO sky density for $D>0.9$~km (upper row) and $D>0.65$~km (lower row) \edit{1}{shown in Figure~\ref{fig:lim}}. The left column shows the \edit{1}{searched} $i'$ and simultaneously observed $g'$ light \edit{1}{curves} for each star with the corresponding templates in black, while the bolded points indicate the corresponding $i'$ and $g'$ inset frames centered on the respective stars shown to the right (1~px = $0''\llap{.}6$). Templates are well-matched to the actually fitted $i'$ light curves \edit{1}{as expected}, but \edit{1}{are} not as well-matched to the \edit{1}{simultaneous} $g'$ light curves, while the star-centered insets do not obviously show changes in the brightness of those stars (although we caution even real fluctuations with similar $\sim$10--20\% amplitude may be only marginally discernible visually), suggesting these events in the $i'$ light curve likely represent artifacts of dense-field photometry rather than true occultations.}
\label{fig:lev}
\end{figure*}

We now look more closely at a few of the \edit{2}{occultation-like signatures that set the upper limits on occultation rate in order} \edit{1}{to investigate \edit{2}{their viability as occultations and} the potential for future improvement \edit{2}{of} these limits}. Figure~\ref{fig:lev} presents two examples of such events setting the $D>0.9$~km and $D>0.65$~km upper limits, showing both their fitted red/$i'$ and simultaneous blue/$g'$ light curves, along with the corresponding templates they matched to, and insets centered on the respective stars for a few key frames around each event.

For all of these examples, the red/$i'$ light curve signatures closely resemble the corresponding templates, thus validating the efficacy of our template matching approach which identified these as red/$i'$ light curves sections with the highest S/N for which these particular occultations cannot be rejected. Note, however, that these templates are not actually the best fit templates to these light curves; lower amplitude occultations actually match these light curves with lower residuals, but give lower template S/N, so \edit{2}{they} do not provide as limiting a constraint as these worse-fitting but higher S/N templates representing occultations that would still produce a light curve with a comparable or worse fit to the template $\beta\sim15.9\%$ of the time.

\edit{1}{Additionally, the simultaneous blue/$g'$ light curves, which were not template-matched in our current analysis (thus serving as a partially independent source of validation), do not appear as well-matched as the red/$i'$ light curves to their corresponding templates.} The star-centered insets also do not reveal any clear dimming corresponding to the dips in the light curves, and instead show slight changes in the PSF around those times. These factors suggest that the occultation-like signatures in the light curves are likely artifacts arising from dense-field photometry errors rather than true variations in the brightness of those stars, so we conclude our limiting events are highly unlikely to be true occultations.

From this qualitative analysis, one potential future improvement to sensitivity could come from jointly fitting occultation templates against both the red/$i'$ and the blue/$g'$ light curves (instead of just red/$i'$, as in our current analysis) to generate a combined template--light curve distance using all of the data. However, the red color of most stars in M22 along with the worse atmospheric turbulence and instrumental image quality in the blue/$g'$ channel led most of the blue/$g'$ light curves to be significantly lower in S/N than the red/$i'$ counterparts. Given also the noticeable correlation between the light curves from the two channels, this effort may only minimally improve our limits with the existing M22 dataset.

The quality of the photometry leaves more clear room for improvement, with the light curves of most stars having S/N well below both the shot noise and atmospheric scintillation limit. About 5,000 stars in the M22 field should have shot noise-limited S/N $>$ 10---far above the typical $\sim$800 light curves with S/N (per correlated length) $>$ 10 indicated in Figure~\ref{fig:fnf}. Meanwhile, atmospheric scintillation alone should affect stars of all brightness similarly, indicating it contributes noise only at 1--2\% of the total flux, given the S/N of the brightest stars. Improving the photometry to these theoretical limits would vastly improve our sensitivity, especially to the smallest, $D<1$~km KBOs whose occultations can only be distinguished in very high S/N light curves.

The extremely high star density in our M22 field, however, presents a serious challenge for any significant photometry improvements, which will also likely be very computationally expensive. \edit{1}{Some improvement to our upper limit may be feasible by reprocessing only the light curves with limiting events and \edit{2}{other occultation-like signatures of similar S/N} using proper PSF fitting methods, where the improved photometry could potentially reject those \edit{2}{features as} being \edit{2}{from} occultations. Due to the spatial variation in PSF across the field, such methods aiming to improve upon our existing photometry must be able to derive a PSF entirely from a dense field without any isolated stars much brighter than surrounding stars, as the few such stars present in this field (used in our method to compute our per-frame ePSFs) are largely distributed toward the outer edge where the PSF may significantly differ from that in the dense core of the cluster.} Any future, follow up survey following our strategy may benefit from a lower density field, which circumvents such complications by allowing large photometric apertures insensitive to small changes in the PSF, which we anticipate \edit{1}{can} more than offset the reduced star count.

\section{Conclusions}

In summary, we used the CHIMERA instrument on the Palomar Hale Telescope to observe stars near the core of the M22 globular cluster for a cumulative $\sim$63~hr over 24 nights to monitor for serendipitous occultations by kilometer-scale KBOs of those stars. We developed and used a modified aperture photometry technique to efficiently extract light curves from the dense star field, then compared the processed light curves against templates of light curve signatures from occultations by kilometer-scale KBOs to set upper limits on the sky density of such KBOs. Finally, we \edit{1}{presented} a couple of examples of limiting events---the highest S/N light curves from which we cannot exclude an occultation---\edit{1}{which appear} to likely represent artifacts arising from our imperfect dense-field photometry procedure rather than actual occultations.

The 95\% confidence upper limits on kilometer-scale KBO abundance set by our nondetections are comparable in magnitude to those set by several other recent occultation surveys, at ${\lesssim}10^7$~deg$^{-2}$ of KBOs $>$1~km in diameter. This limit, however, remains insufficiently constraining for comparison with several of the claimed occultations, which will require at least another order-of-magnitude improvement in sensitivity to robustly validate or reject. Future surveys following a similar strategy may find such an improvement in observing efficiency by targeting lower density star fields with well-separated stars that permit high quality aperture photometry, avoiding the dense-field photometry artifacts that limited the S/N of our light curves to well below theoretical scintillation and shot noise limits. Such improvements, together with the addition \edit{1}{of} a high latitude control data set constraining the baseline noise, may eventually permit constraints on not only the number density but also orbital distribution of these kilometer-sized KBOs once a sample of true occultations can be definitively isolated.

\bigskip

This work was supported by a grant from the Simons Foundation (668346, JPG). H.E.S. gratefully acknowledges NASA grant 80NSSC18K0828 for financial support during preparation and submission of the work.

This research has made use of observations from the Hale Telescope at Palomar Observatory, which is owned and operated by Caltech and administered by Caltech Optical Observatories.

We also thank the reviewer for their helpful comments and suggestions.

\facilities{Hale (CHIMERA)}
\software{Astropy \citep{astropy2022}, Matplotlib \citep{hunter2007}, NumPy \citep{vanderwalt2011}, SciPy \citep{virtanen2020}}

\bibliography{ms}

\end{document}